\documentclass[prl,aps,twocolumn,amsmath,amssymb,superscriptaddress]{revtex4-1}

\usepackage{amsmath}  
\usepackage{amsthm}  
\usepackage{amssymb}	 
\usepackage{graphicx}  
\usepackage[usenames,dvipsnames]{color}
\usepackage{array} 
\usepackage{paralist} 
\usepackage{amsfonts}
\usepackage{mathtools}
\usepackage{times}
\usepackage{braket}
\usepackage{slashed}
\usepackage[colorlinks=true,linkcolor=blue,citecolor=blue,breaklinks]{hyperref}
\normalfont


\renewcommand{\vec}[1]{\ensuremath{\mathbf{#1}}} 
 
\newcommand{\avg}[1]{\left< #1 \right>} 
 
\let\baraccent=\= 
\renewcommand{\=}[1]{\stackrel{#1}{=}} 

\theoremstyle{definition}

\theoremstyle{remark}

\newcommand{\be}{\begin{equation}}
\newcommand{\ee}{\end{equation}}

\newcommand{\bea}{\begin{eqnarray}}
\newcommand{\eea}{\end{eqnarray}}
\newcommand{\beal}{\begin{align}}
\newcommand{\eeal}{\end{align}}

\begin{document}

\title{Field-Driven Gapless Spin Liquid in the Spin-1 Kitaev Honeycomb Model}
\author{Ciar\'{a}n Hickey}
\email[E-mail: ]{chickey@thp.uni-koeln.de}
\affiliation{Institute for Theoretical Physics, University of Cologne, 50937 Cologne, Germany}
\author{Christoph Berke}
\affiliation{Institute for Theoretical Physics, University of Cologne, 50937 Cologne, Germany}
\author{Panagiotis Peter Stavropoulos}
\affiliation{Department of Physics, University of Toronto, Ontario M5S 1A7, Canada}
\author{Hae-Young Kee}
\affiliation{Department of Physics, University of Toronto, Ontario M5S 1A7, Canada}
\affiliation{Canadian Institute for Advanced Research, CIFAR Program in Quantum Materials, Toronto, ON M5G 1M1, Canada}
\author{Simon Trebst}
\affiliation{Institute for Theoretical Physics, University of Cologne, 50937 Cologne, Germany}

\begin{abstract}
Recent proposals for spin-1 Kitaev materials, such as honeycomb Ni oxides with heavy elements of Bi and Sb, have shown that these compounds naturally give rise to {\em antiferromagnetic} (AFM) Kitaev couplings. Conceptual interest in such AFM Kitaev systems has been sparked by the observation of a transition to a gapless $U(1)$ spin liquid at intermediate field strengths in the AFM \mbox{spin-1/2} Kitaev model. 
However, all hitherto known \mbox{spin-1/2} Kitaev materials exhibit ferromagnetic bond-directional exchanges. 
Here we discuss the physics of the spin-1 Kitaev model in a magnetic field and show, by extensive numerical analysis, that for AFM couplings it exhibits an extended gapless quantum spin liquid at intermediate field strengths. The close analogy to its \mbox{spin-1/2} counterpart suggests that this gapless spin liquid is a $U(1)$ spin liquid with a neutral Fermi surface, that gives rise to enhanced thermal transport signatures. 
\end{abstract}
\maketitle


Quantum spin liquids are disordered phases of matter that exhibit fractionalization of the underlying spin degrees of freedom and an associated emergent gauge structure \cite{Savary2017quantum, zhou_quantum_2017, broholm_quantum_2020, balents_spin_2010}. These properties manifest in various guises, with a rich variety of quantum spin liquid (QSL) states theoretically studied so far, and a number of associated material candidates identified \cite{trebst_kitaev_2017,knolle_field_2019,Stavropoulos_S1}. A powerful framework to describe these states is the idea of splitting the fundamental spin degrees of freedom into constituent partons, either fermions or bosons. These describe the low-energy fractionalized quasiparticles of the QSL state, and are naturally accompanied by a gauge structure that enforce local constraints. However, making concrete connections between these theoretical concepts, candidate materials and experimental signatures remains a challenge \cite{knolle_field_2019}. 

One of the most well-studied models for a QSL is Kitaev's honeycomb model \cite{Kitaev2006anyons}. This \mbox{spin-1/2} model, consisting of bond-dependent Ising interactions, is in fact exactly solvable. Using fermionic partons, the ground state corresponds to a nodal superconductor (SC) coupled to a static $\mathbb{Z}_2$ gauge field -- a gapless $\mathbb{Z}_2$ QSL \cite{NayakParton2011}. Dominant Kitaev interactions can actually be realized, given the right ingredients, in a number of spin-orbit entangled $j=1/2$ Mott insulators \cite{Jackeli2009}. These Kitaev materials have been shown to display many properties similar to those theoretically predicted for the Kitaev model, despite the inevitable presence of additional non-Kitaev interactions that drive magnetic order at the lowest temperatures \cite{Plumb2014spin,Banerjee2016proximate,Banerjee2016neutron,BaenitzRuClMF2015,Sears_magnetic_order_2015,Singh_antiferromagnetic_mott_2010,chun_direct_2015,revelli2019fingerprints}. Though there is ongoing debate over the relevant form and magnitude of such non-Kitaev interactions, it is generally accepted that the dominant Kitaev interaction is ferromagnetic (FM) in nature \cite{WinterValenti2016, Winter_2017}.

The nodal structure of the Kitaev QSL can be gapped out by applying an infinitesimal magnetic field, producing a non-trivial topological SC of fermionic partons \cite{NayakParton2011}. This state contains non-Abelian anyon excitations \cite{Nayak_nonabelian_2008}, the vortex excitations of the SC, and a chiral Majorana edge state, which gives a half-quantized thermal Hall conductance \cite{Kitaev2006anyons}. In the case of FM Kitaev interactions, relevant for the traditional set of $j=1/2$ candidate materials, further increasing the magnetic field quickly destroys the topological state and leads directly to the trivial partially polarized state \cite{Jiang2011possible}. On the other hand, for antiferromagnetic (AFM) interactions, increasing the field leads to a transition (at considerably higher magnetic field strength \cite{FeyMSc2013,ZhuAFMK2017}) to a gapless QSL with an emergent $U(1)$ gauge structure \cite{Hickey2019,PollmannAFMK2018,jiang2018field,Patel2019,Nasu2018,Kaib2019}, akin to 2+1d quantum electrodynamics. In the parton framework, this corresponds to a SC to metal transition, resulting in the unusual situation of a magnetic insulator with an emergent Fermi surface. The possibility of realizing such an exotic state motivates the search for new Kitaev material candidates that naturally realize dominant AFM, rather than FM, Kitaev interactions.  

A recent theoretical study \cite{Stavropoulos_S1} suggested a new class of Kitaev materials that realize a spin-1 version of the Kitaev honeycomb model, with a number of specific material candidates proposed. Crucially, the microscopic mechanism responsible naturally leads to an AFM sign of the bond-directional interaction. The necessary ingredients include strong Hund's coupling between cation $e_g$ electrons and strong spin-orbit coupling for the anion electrons. Examples include honeycomb Ni oxides such as $A_3$Ni$_2 X$O$_6$ with $A$ = Na, Li, and $X$ = Bi, Sb. Notably, while they exhibit zigzag magnetic ordering at low temperature, specific heat measurements show an entropy plateau of $1/2 \log 3$ well above the N\'eel temperature \cite{zvereva_zigzag_2015}, suggesting strong fluctuations indicating proximity to Kitaev dominated physics. Though the spin-1 Kitaev model, relevant to these materials, is not exactly solvable, it shares many of the same properties and phenomenology of the \mbox{spin-1/2} version \cite{baskaran_orderdisorder_2008}. This raises the immediate question of how the spin-1 model behaves in the presence of a magnetic field and specifically whether a gapless $U(1)$ QSL appears, that is thus amenable to physical realization in this new class of materials.

Here, we study the spin-1 Kitaev model in an external magnetic field using numerical exact diagonalization techniques \cite{sandvikreview,arpackusers}. By computing a range of static, dynamical and finite temperature properties we show that the field-driven physics of the spin-1 model bears many striking similarities to the \mbox{spin-1/2} case. For example there is a clear intermediate field regime with a dense continuum of states carrying featureless spin spectral weight at very low energies. We argue that the \mbox{spin-1} model indeed realizes a gapless $U(1)$ QSL at intermediate fields, and comment on its stability and experimental fingerprints in material candidates. We briefly remark on the low-field regime and the potential nature of a $\mathbb{Z}_2$  QSL there.
Finally, we investigate the effects of adding a finite FM Heisenberg interaction, the most relevant additional term produced by the microscopic mechanism of Ref. \cite{Stavropoulos_S1}. 

\vspace{0.05cm}


\noindent {\it Model.--}
The Kitaev honeycomb model consists of bond-dependent Ising interactions between local moments. Here, we are interested in the case of spin-1 local moments, with the field dependent Hamiltonian given by  
\begin{equation}
H =  K \sum_{\avg{i,j}\in\gamma} S_i^\gamma S_j^\gamma - \sum_i \vec{h} \cdot \vec{S}_i \,,
\end{equation}
where $K>0$ is an AFM Kitaev coupling and the bond directions are denoted by $\gamma\in \left\{ x,y,z \right\}$. The magnetic field can in principle point in any direction. Here, we focus on a generic direction away from any special high-symmetry lines, specifically a field tilted $\pi/24$ away from the $[111]$ direction toward the $[1\bar{1}0]$ direction, ensuring that $h_x\neq h_y \neq h_z$ \footnote{{Note that we are using a normalized magnetic field vector, e.g.~a field along the $[111]$ direction would be $\vec{h}=\left(h,h,h\right)/\sqrt{3}$}} (tilting the field further does not result in qualitative changes to the results presented here). For material candidates, the $[111]$ direction naturally corresponds to the out-of plane $c$-axis. 

In the absence of an external field, the model has an extensive number of conserved quantities. These are given by the plaquette operators $W_p = \exp\left[ i \pi \left( S_i^x + S_j^y + S_k^z + S_l^x + S_m^y + S_n^z \right)\right] $, where $S_i^\alpha$ is the spin at site $i$ with $\alpha$ the bond not included in the plaquette and with $\mathbb{Z}_2$ eigenvalues $\avg{W_p} = \pm 1$. Though this is not enough to ensure an exact solution, as in the \mbox{spin-1/2} case, it does guarantee that spin-spin correlation functions precisely vanish beyond nearest-neighbor \cite{baskaran_orderdisorder_2008}. There is thus no possibility of conventional long-range magnetic order. Furthermore, applying a spin operator $S_i^\alpha$ at site $i$ flips the sign of $\avg{W_p}$ on the two plaquettes neighboring the $\alpha$-bond. Numerically, the ground state is found in the sector $\avg{W_p} = + 1, \forall p$ \cite{Koga_Spin1}, as in the \mbox{spin-1/2} case. The spin operator $S_i^\alpha$ thus creates two plaquette excitations, i.e. two plaquettes with $\avg{W_p} \neq + 1$. 

\vspace{0.05cm}

\begin{figure}[!t]  
\includegraphics[width=\columnwidth]{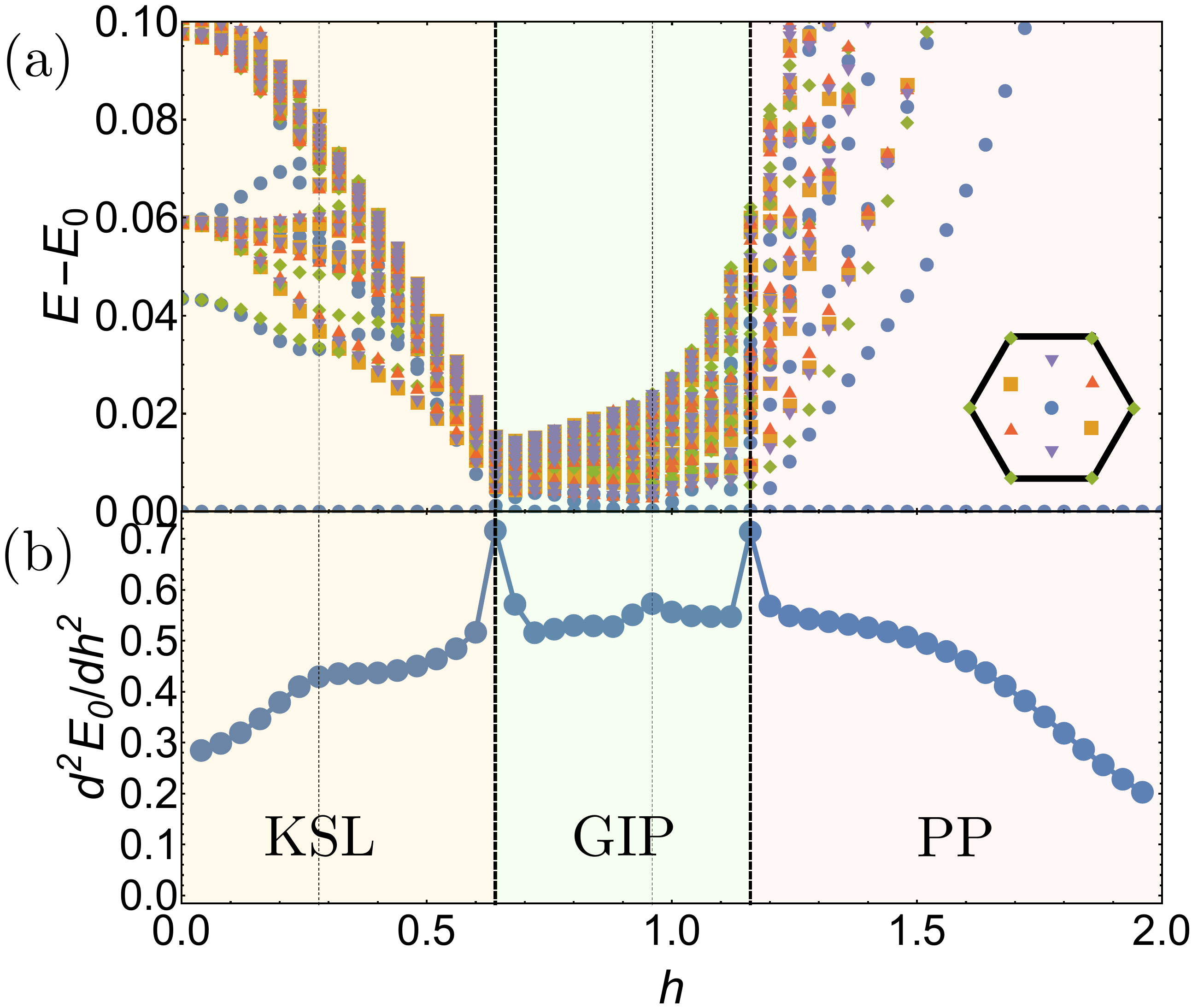}
\caption{(a) {\bf Energy spectrum} as a function of field for the $N=18$ site symmetric cluster. For each field value the ten lowest energy states in each momentum sector are shown. (b) {\bf Second derivative of the ground state energy} as a function of field. The dotted lines indicate peaks in the second derivative with thick (thin) lines indicating strong (weak) features.}
\label{fig:En}
\end{figure}

\noindent {\it Phase diagram in field--}
The general form of the in-field phase diagram can be mapped out by identifying phase boundaries through scans of the energy  spectrum of the above Hamiltonian as a function of field magnitude $h$, as well as the second derivative of the ground state energy $d^2 E_{0} /dh^2$. Results from such scans are shown in Fig.~\ref{fig:En} for an $N=18$ site cluster (which maintains all of the symmetries of the honeycomb lattice). There are three distinct regions clearly visible, separated by sharp signatures in $E_{0}$. The high-field region, $h>1.16$, is easily identifiable as the partially polarized (PP) state, smoothly connected to a trivial fully polarized product state as $h\rightarrow \infty$. The spectrum in the low-field region, $h<0.64$, is structurally similar to the zero-field limit. We thus refer to this region as the Kitaev spin liquid (KSL) region, and will discuss it in more detail later. Finally, the intermediate region, $0.64<h<1.16$, presents a new phase, fundamentally distinct from either of the limits $h\rightarrow 0$ or $h\rightarrow \infty$. It is clearly marked by a dense continuum of low-energy states, e.g.~for the $N=18$ site cluster used here there are $\sim \!\! 100$ states with energies $E-E_0 < 0.02 K$. This remarkable low-energy density of states strongly suggests the presence of gapless excitations, and we thus label it as the gapless intermediate phase (GIP). Due to the high computational cost of diagonalization, it is unfortunately not possible to carry out a thorough finite-size scaling analysis of this region.  

\begin{figure}[!t]  
\includegraphics[width=\columnwidth]{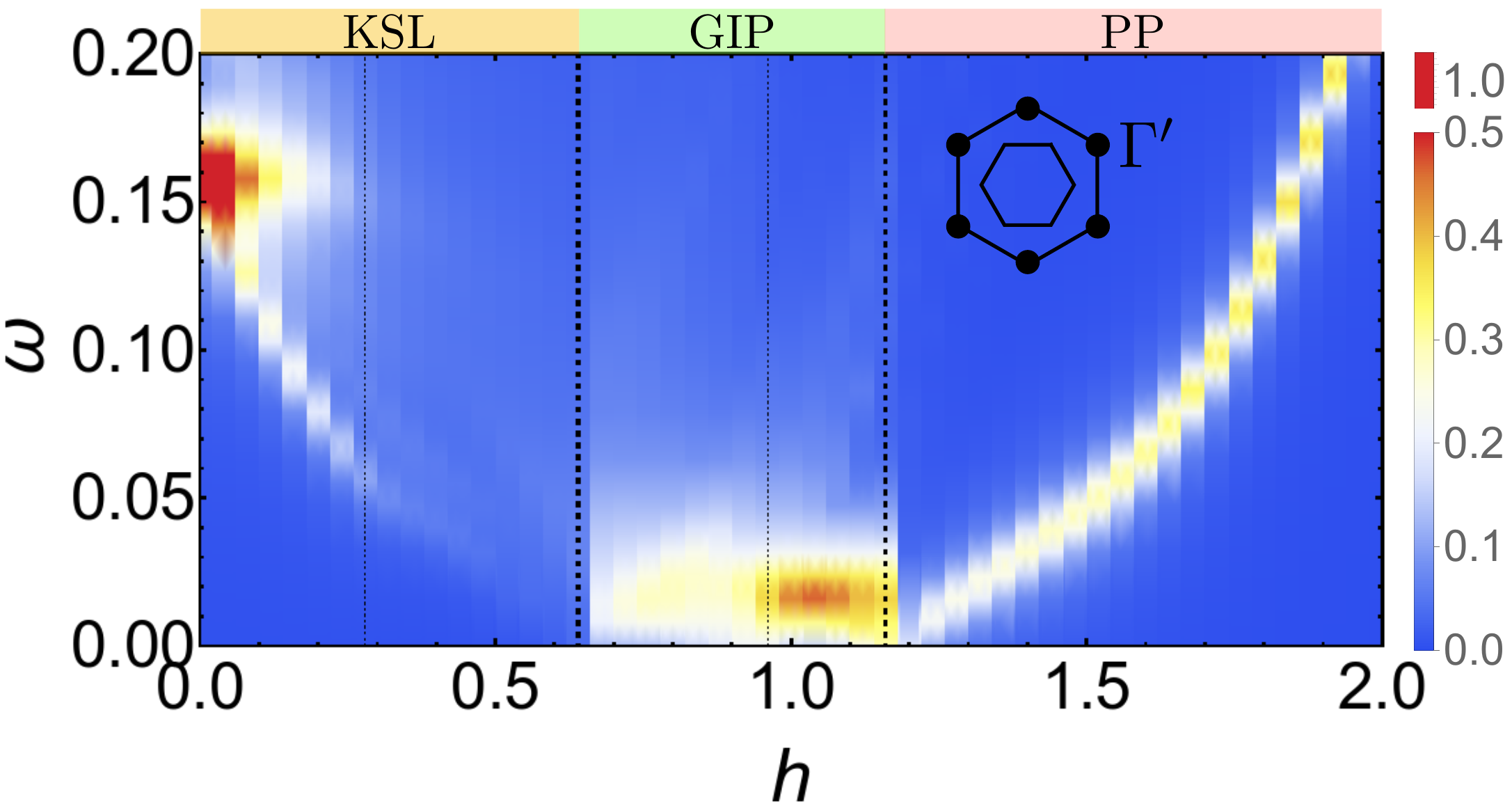}
\caption{{\bf Dynamical spin structure factor} at the $\Gamma^\prime$ point, $S\left(\Gamma^\prime,\omega\right)$. At low-fields there is a clear gap, which vanishes on entry to the intermediate phase, and then re-opens in the polarized phase with a single sharp mode present, corresponding to the lowest magnon excitation.}
\label{fig:DSF}
\end{figure}

\begin{figure}[!t]  
\includegraphics[width=\columnwidth]{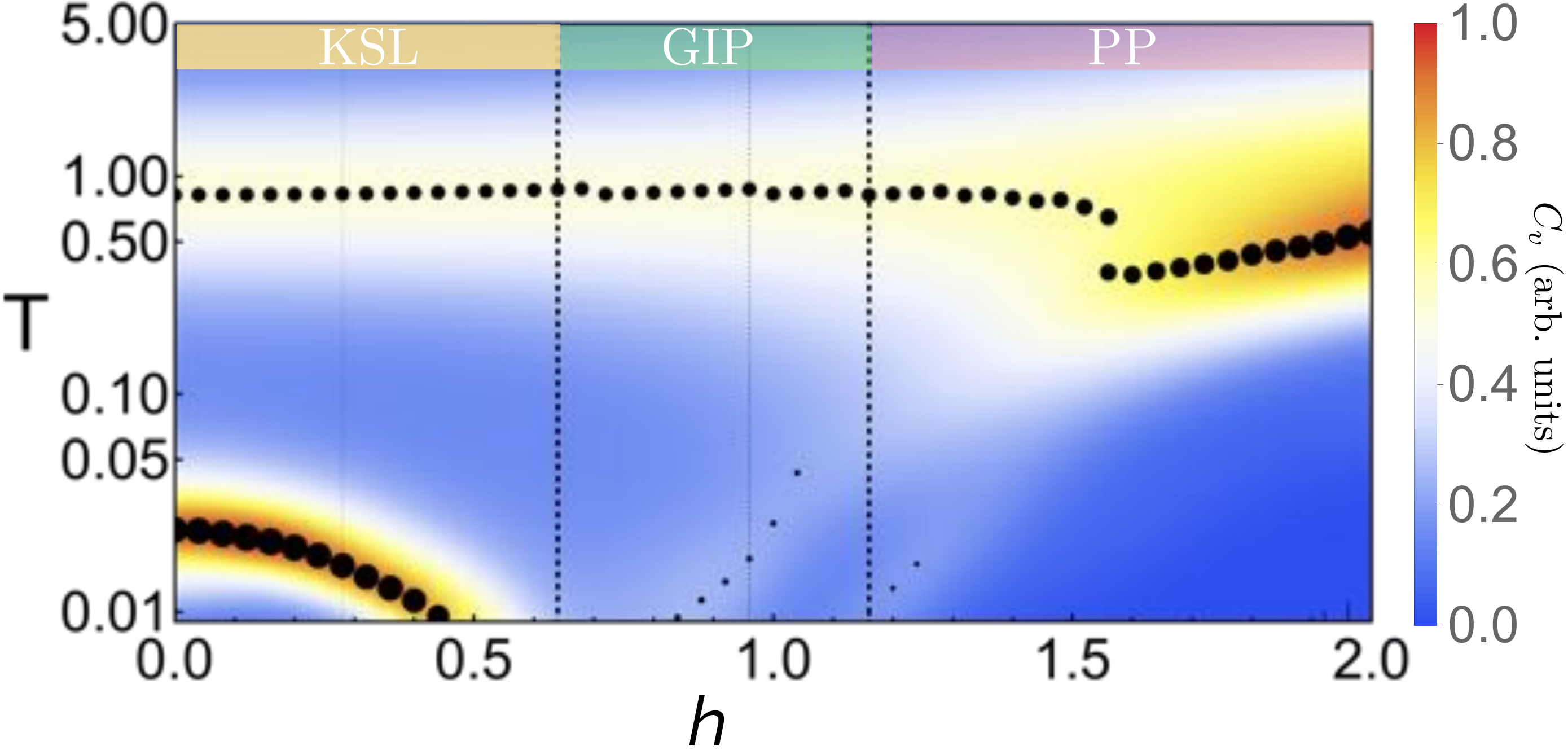}
\caption{{\bf Specific heat} $C_v(T)$ as a function of increasing field. At zero-field, there is a two-peak structure, with the lower-temperature peak, associated with the plaquette variables, sharply bending downwards as the field increases, disappearing below the temperature range accessible in our calculations in the intermediate phase. At high-fields, there is a single peak, as expected. The black dots indicate peaks in the data, with their size scaled relative to the peak height.}
\label{fig:Cv}
\end{figure}

The phase diagram and spectrum bear remarkable resemblance to the \mbox{spin-1/2} case. Indeed, scaling the field magnitude by the spin length $S$, the upper and lower critical fields for the intermediate phase are $h/S=0.85, 1.18$ for the \mbox{spin-1/2} case, and $h/S=0.64, 1.16$ for the \mbox{spin-1} case (for the same $N=18$ site cluster and field angle). However, we note that in the \mbox{spin-1} case there are two additional weak features in $d^2 E_{0} /dh^2$, shown in Fig.~\ref{fig:En}(b), at $h=0.28$ and $h=0.96$. These fields do not correspond to any qualitative changes in other physical quantities (as evidenced in Figs.~\ref{fig:DSF} and \ref{fig:Cv}), unlike the sharp peaks discussed above, and thus may simply be a result of finite-size effects, though this requires further study.  

Note that for a FM Kitaev coupling the structure of the phase diagram of the \mbox{spin-1}  model is identical to that of the \mbox{spin-1/2} model \cite{ZhuAFMK2017,Hickey2019}. There is just a single phase transition as a function of field, from the low-field KSL to the high-field PP phase, at a critical field of $h=0.04$, an order of magnitude smaller than the AFM case. This dramatic disparity in critical fields is, as in the \mbox{spin-1/2} case \cite{Hickey2019}, a direct consequence of the different sign of the ground state spin-spin correlations, FM vs AFM, with the FM ground state being much more susceptible to polarization by a uniform magnetic field.

\vspace{0.05cm}


\noindent {\it Intermediate phase.--} 
The nature of the intermediate phase, and its similarities to the \mbox{spin-1/2} case, can be readily seen in its dynamical and finite temperature properties. In Fig.~\ref{fig:DSF} we show the dynamical spin structure factor at the $\Gamma^\prime$ point (we plot only the diagonal contributions $ \sum_\alpha S^{\alpha\alpha} \left( \Gamma^\prime, \omega\right) $). At zero field, there is a clear gap in the dynamical structure factor, with a strong feature at $\omega=0.16$. This is true despite an abundance of states within the gap, they simply carry zero spin spectral weight and are thus invisible in such a plot. As the structure factor measures the response to single spin excitations $S_i^\alpha$ which, as discussed above, correspond to plaquette excitations, this indicates that plaquette excitations are gapped at zero-field. As the field is increased, the gap gets smaller and smaller, and the intensity of the feature significantly decreases, until it reaches zero energy at the transition to the intermediate phase. In the intermediate phase itself  the spin spectral weight remains concentrated at low energies, spread between the dense continuum of low-lying states seen in Fig.~\ref{fig:En}(a). This behavior is almost identical to the \mbox{spin-1/2} case, in which the gap represents the energy scale for creating a pair of flux excitations of a $\mathbb{Z}_2$ gauge field \cite{Knolle_Dynamics_2014}. There, the collapse of the gap has been interpreted as marking the transition from a gapped to a gapless gauge field, suggesting a similar scenario here. One significant quantitative difference between the \mbox{spin-1/2} and \mbox{spin-1} cases is that, in the KSL, the spectral weight at the gap decreases considerably faster in the spin-1 case as field is increased, whereas the ratio of the integrated intensity at low energies in the intermediate phase to the integrated intensity of the zero-field feature is roughly the same, $\sim 40\%$, in both cases \cite{Hickey2019}. 

Next we turn to thermodynamic signatures, with the specific heat as a function of field magnitude shown in Fig.~\ref{fig:Cv}, calculated using the method of thermal pure quantum states \cite{TPQS2012,CanTPQS2013}. There are again three distinct regions easily visible, corresponding to the KSL, intermediate phase and PP phase. At zero field there is a two-peak structure to the specific heat \cite{oitmaa_2018}, just as in the \mbox{spin-1/2} case \cite{Nasu2014vaporization}. This reflects the two defining properties of the ground state, finite expectation values for purely nearest-neighbor spin-spin correlations and for the plaquette operators. For the \mbox{spin-1/2} case, these properties reflect fractionalization of the spin degrees of freedom, and are associated with the mobile Majorana fermions and static flux excitations, respectively. At the higher-$T$ peak, the spin-spin correlations reach their zero-$T$ values whereas at the lower-$T$ peak the plaquette operators saturate to their zero-T values.    

As the magnetic field is increased we see that the lower-$T$ peak strongly bends to lower temperatures, while the higher-$T$ peak remains virtually unchanged. This indicates that the transition is driven by the plaquette degrees of freedom, with the drop of this energy scale consistent with the drop of the gap in the dynamical spin structure factor. In the high-field PP phase a single peak at $T\sim \mathcal{O}(K)$ is recovered, as expected. This behavior is again essentially identical to the \mbox{spin-1/2} case.   

\begin{figure}[!t]  
\includegraphics[width=\columnwidth]{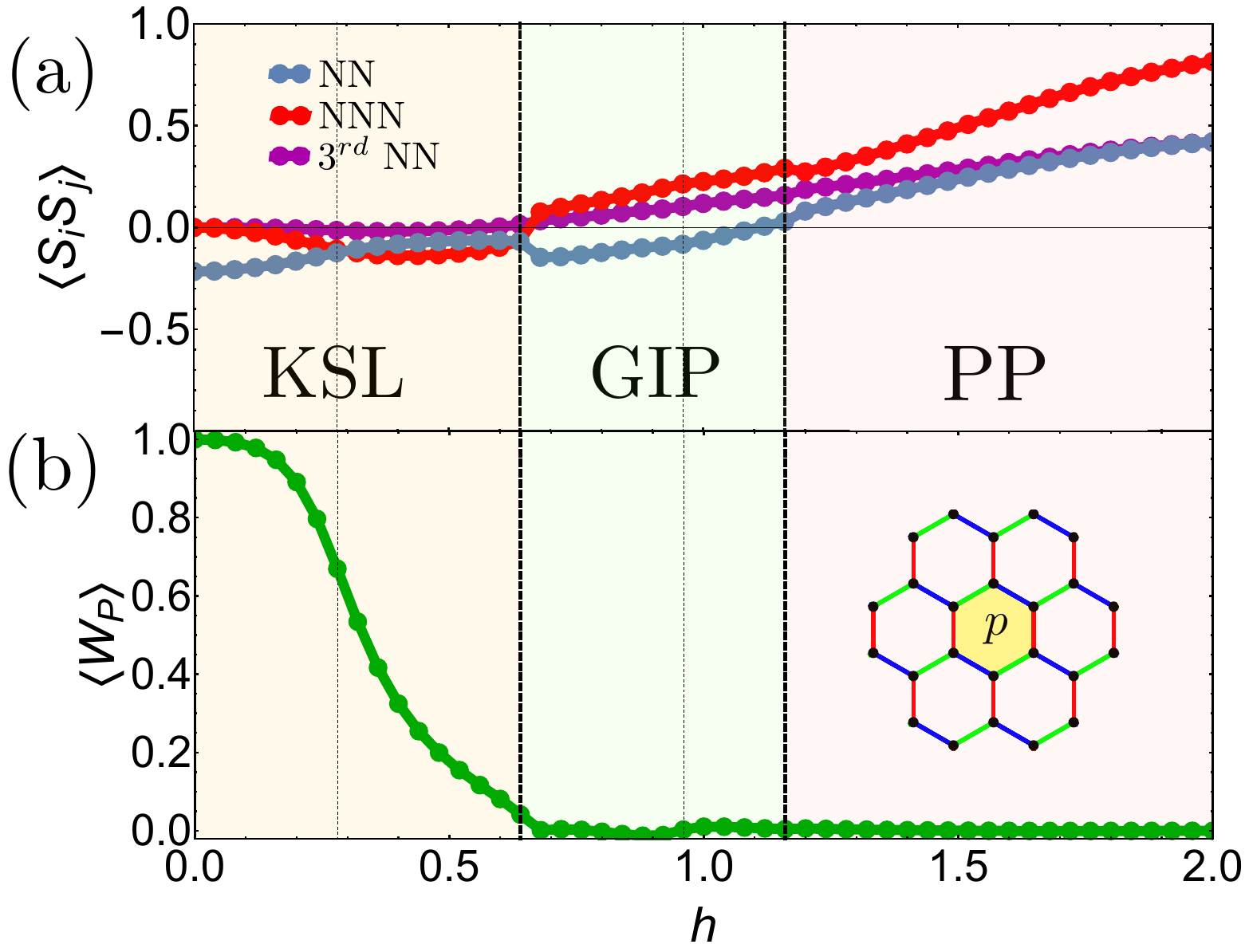}
\caption{(a) {\bf Spin-spin correlations} for nearest-neighbor (NN), next-nearest-neighbor (NNN) and third-nearest-neighbor ($3^{rd}$ NN) as a function of increasing field. (b) {\bf Expectation value of the plaquette operator} $\avg{W_p}$, which vanishes once the transition to the intermediate phase is crossed. }
\label{fig:Static}
\end{figure}

Finally, we remark on some static ground state properties of the intermediate phase. In Fig.~\ref{fig:Static}(a), we show the nearest-neighbor, next-nearest-neighbor, and third-nearest-neighbor spin-spin correlation functions as a function of magnetic field. The low-field KSL state is characterized by AFM, i.e. negative, spin-spin correlations, with the transition to the intermediate phase accompanied by a change in sign of the second- and third-nearest neighbor correlations, and then the final transition to the PP state resulting when all correlations become FM, i.e. positive. In Fig.~\ref{fig:Static}(b), we plot the expectation value of the plaquette operator $\avg{W_p}$. While we have $\avg{W_p}=+1$ at zero field, this quickly decreases and goes to zero at the transition to the intermediate phase. It remains zero throughout the intermediate phase and does not display any signatures as the transition to the PP is crossed. In this case, this behavior is actually in marked contrast to the \mbox{spin-1/2} case, where $\avg{W_p}$ continuously varies throughout the intermediate phase, only reaching zero at the transition to the trivial PP phase.

\vspace{0.05cm}

The microscopic mechanism responsible for the generation of the spin-1 Kitaev model in candidate materials also generates a finite FM Heisenberg interaction. Interestingly, a symmetric off-diagonal exchange term, $\Gamma$, is absent in the fourth order perturbation analysis of Ref. \cite{Stavropoulos_S1}. We thus consider only the effects of a FM Heisenberg term.

In the absence of a field the KSL is destroyed when $J/K = 0.08$. The phase becomes less robust as field is increased, with the phase boundary shifting to smaller and smaller $J/K$. On the other hand, the intermediate phase appears to cover a much smaller region of parameter space, with a maximum extent of at most $J/K\sim 0.01$. This demonstrates that, though the intermediate phase is indeed stable to the physically relevant addition of a finite FM Heisenberg term, it is significantly less stable than the KSL. We also note that if a material is already magnetically ordered at zero field, it is not possible to access either the KSL or the intermediate phase at finite field (at least within the Kitaev-Heisenberg model considered here).

\vspace{0.05cm}

%
%


\noindent{\it Discussion.--}
Our numerical analysis of the spin-1 AFM Kitaev model in the presence of a uniform magnetic field using exact diagonalization has yielded a phase diagram, where the evolution of energy spectrum, dynamical structure factor and specific heat as a function of field are 
all qualitatively similar to the corresponding \mbox{spin-1/2} case. 
These results might perhaps not come completely unexpected. Firstly, though increasing the spin magnitude from $S=1/2$ to $S=1$ generically reduces quantum fluctuations, it need not necessarily significantly alter the physics of a system, particularly in highly frustrated models such as the Kitaev model, which does not exhibit long-range magnetic order even in the classical $S\rightarrow\infty$ limit \cite{Chandra2010,Sela2014}. Secondly, though the zero-field spin-1 model is not exactly solvable, as it is in the \mbox{spin-1/2} case, it does share a number of its characteristic features, including an extensive number of conserved $\mathbb{Z}_2$ plaquette variables, purely nearest neighbor spin-spin correlations, a two-peak structure in specific heat and a sharp gap in the dynamical structure factor \cite{baskaran_orderdisorder_2008,Koga_Spin1,oitmaa_2018}. Taken together, our numerical results thus imply that the nature of the gapless phase that appears at intermediate field strengths is qualitatively the same in both the spin-1 and \mbox{spin-1/2} cases. This leads us to conclude that the intermediate phase is a $U(1)$ quantum spin liquid with a neutral Fermi surface. 


This characterization of the intermediate phase also allows us to shed some light on the nature of the low-field phase. In the language of fermionic partons, the intermediate phase would correspond to a metallic phase, with a Fermi surface of partons coupled to an emergent $U(1)$ gauge field. One possible scenario for the transition into the low-field phase, which has first been put forward in the case of spin-1/2 moments \cite{Hickey2019}, is that this transition is a metal to superconductor transition. For such an Anderson-Higgs transition, it is the $U(1)$ gauge field that would be Higgsed, i.e.\ become massive, and turn into a $\mathbb{Z}_2$ gauge field. Translating back to spin language, this would be a transition connecting two QSLs -- a $U(1)$ QSL with a $\mathbb{Z}_2$ QSL. Such a scenario is particularly tempting if one recalls that the spin-1 Kitaev model at zero-field indeed possesses an extensive number of conserved $\mathbb{Z}_2$ plaquette variables, which further suggests an emergent $\mathbb{Z}_2$ gauge structure \footnote{{At the very least one would expect a QSL with a finite gauge group $\mathbb{Z}_N$ (so that gauge excitations remain gapped). The possibility of realizing a scenario with $N>2$, due to the larger Hilbert space of the higher-spin Kitaev models, is an interesting question.}}.
Going one step further, it is then natural to ask whether this low-field phase is, in the language of the partons, a gapless or a gapped SC. Maintaining a gapless SC throughout an extended region of parameter space generally requires either fine-tuning or the existence of symmetries to protect it. The model and field direction we consider here possess very little symmetry so it is unlikely a gapless SC could be stabilized. In the case of a gapped SC, it can either possess an odd Chern number, as in the \mbox{spin-1/2} case ($C=\pm1$), or an even Chern number (including $C=0$). However, distinguishing between these two possibilities, which in spin language correspond to non-Abelian and Abelian QSLs, respectively, is a major challenge. On a more speculative note, one interesting possibility is that the half-integer KSLs realize non-Abelian QSLs and the integer KSLs Abelian ones, or in parton language the KSL is a SC with, e.g.~a Chern number $C=2S$.       

Finally we comment on the potential experimental relevance and observable signatures of the physics discussed here. 
It is important to note that a crucial ingredient are {\em antiferromagnetic} Kitaev interactions.
While such antiferromagnetic couplings have not been observed for the current family of spin-1/2 Kitaev materials \cite{trebst_kitaev_2017}, it has
been argued that they arise naturally for spin-1 Kitaev materials \cite{Stavropoulos_S1}.
A number of candidate compounds for such spin-1 Kitaev materials have been put forward, with honeycomb Ni oxides such as $A_3$Ni$_2X$O$_6$, with $A$ = Na, Li, and $X$ = Bi, Sb, being amongst the most promising \cite{Stavropoulos_S1}. Assuming that a candidate material with AFM Kitaev interactions \footnote{{Note that there have also been alternative routes to realizing AFM Kitaev interactions proposed, with spin-1/2 moments \cite{motome_AFM_fbased,sugita2019} }} is found, what are the expected observable signatures of the intermediate phase? For the gapless QSL with a neutral Fermi surface put forward in our line of arguments, one expects, e.g.~lack of long-range order, power law heat capacity, finite thermal conductivity as $T\rightarrow 0$ and a featureless inelastic neutron scattering response with spin spectral weight at low energies across the Brillouin zone. However, it may be that a candidate material, due to non-negligible non-Kitaev interactions, does not actually exhibit the phase as a ground state, but instead may lie `proximate' to it. In this case, signatures of the physics here may still be accessible at elevated temperatures and energies, as discussed for the current generation of Kitaev materials \cite{Banerjee2016proximate,Banerjee2016neutron,revelli2019fingerprints}, but concrete predictions for such a `proximate $U(1)$ spin liquid' requires further study.\\

        \textit{Note added:} During completion of this manuscript two related preprints \cite{zhu_magnetic_2020,YB_2020} were posted which use density matrix renormalization group techniques to explore the phase diagram of the spin-1 Kitaev model in a magnetic field. These calculations concentrate mainly on the low-field KSL phase, presenting multifaceted numerical evidence for a gapped $\mathbb{Z}_2$ spin liquid in this regime -- complementing our study focusing on the nature of the intermediate gapless phase. Taken together, these studies provide strong support for the 
scenario of an Anderson-Higgs transition in the spin-1 Kitaev model as put forward in our discussion.

\vskip 1mm

{\it Acknowledgements.} 
We thank A. Laeuchli for useful discussions. 
C.~H. and H.-Y.~K. thank the Aspen Center for Theoretical Physics for hospitality and support under Grant No. NSF 1066293. C.~H., C.~B. and S.~T. acknowledge partial support from the Deutsche Forschungsgemeinschaft (DFG, German Research Foundation), Projektnummer 277101999 and 277146847 -- TRR 183 (project B01) and CRC 1238 (projects C02, C03). 
H.-Y.~K. and P.~P.~S. acknowledge support from the NSERC discovery grant 06089-2016. The numerical simulations were performed on the JUWELS cluster at the Forschungszentrum J\"ulich and the CHEOPS cluster at RRZK Cologne.


\bibliography{KitaevSpin1Field.bib}

\begin{thebibliography}{49}%
\makeatletter
\providecommand \@ifxundefined [1]{%
 \@ifx{#1\undefined}
}%
\providecommand \@ifnum [1]{%
 \ifnum #1\expandafter \@firstoftwo
 \else \expandafter \@secondoftwo
 \fi
}%
\providecommand \@ifx [1]{%
 \ifx #1\expandafter \@firstoftwo
 \else \expandafter \@secondoftwo
 \fi
}%
\providecommand \natexlab [1]{#1}%
\providecommand \enquote  [1]{``#1''}%
\providecommand \bibnamefont  [1]{#1}%
\providecommand \bibfnamefont [1]{#1}%
\providecommand \citenamefont [1]{#1}%
\providecommand \href@noop [0]{\@secondoftwo}%
\providecommand \href [0]{\begingroup \@sanitize@url \@href}%
\providecommand \@href[1]{\@@startlink{#1}\@@href}%
\providecommand \@@href[1]{\endgroup#1\@@endlink}%
\providecommand \@sanitize@url [0]{\catcode `\\12\catcode `\$12\catcode
  `\&12\catcode `\#12\catcode `\^12\catcode `\_12\catcode `\%12\relax}%
\providecommand \@@startlink[1]{}%
\providecommand \@@endlink[0]{}%
\providecommand \url  [0]{\begingroup\@sanitize@url \@url }%
\providecommand \@url [1]{\endgroup\@href {#1}{\urlprefix }}%
\providecommand \urlprefix  [0]{URL }%
\providecommand \Eprint [0]{\href }%
\providecommand \doibase [0]{http://dx.doi.org/}%
\providecommand \selectlanguage [0]{\@gobble}%
\providecommand \bibinfo  [0]{\@secondoftwo}%
\providecommand \bibfield  [0]{\@secondoftwo}%
\providecommand \translation [1]{[#1]}%
\providecommand \BibitemOpen [0]{}%
\providecommand \bibitemStop [0]{}%
\providecommand \bibitemNoStop [0]{.\EOS\space}%
\providecommand \EOS [0]{\spacefactor3000\relax}%
\providecommand \BibitemShut  [1]{\csname bibitem#1\endcsname}%
\let\auto@bib@innerbib\@empty
\bibitem [{\citenamefont {{Savary}}\ and\ \citenamefont
  {{Balents}}(2017)}]{Savary2017quantum}%
  \BibitemOpen
  \bibfield  {author} {\bibinfo {author} {\bibfnamefont {L.}~\bibnamefont
  {{Savary}}}\ and\ \bibinfo {author} {\bibfnamefont {L.}~\bibnamefont
  {{Balents}}},\ }\bibfield  {title} {{\color{Gray}\small \bibinfo {title}
  {{Quantum spin liquids: a review}},\ }}\href {\doibase
  10.1088/0034-4885/80/1/016502} {\bibfield  {journal} {\bibinfo  {journal}
  {Reports on Progress in Physics}\ }\textbf {\bibinfo {volume} {80}},\
  \bibinfo {eid} {016502} (\bibinfo {year} {2017})}\BibitemShut {NoStop}%
\bibitem [{\citenamefont {Zhou}\ \emph {et~al.}(2017)\citenamefont {Zhou},
  \citenamefont {Kanoda},\ and\ \citenamefont {Ng}}]{zhou_quantum_2017}%
  \BibitemOpen
  \bibfield  {author} {\bibinfo {author} {\bibfnamefont {Y.}~\bibnamefont
  {Zhou}}, \bibinfo {author} {\bibfnamefont {K.}~\bibnamefont {Kanoda}}, \ and\
  \bibinfo {author} {\bibfnamefont {T.-K.}\ \bibnamefont {Ng}},\ }\bibfield
  {title} {{\color{Gray}\small \bibinfo {title} {Quantum spin liquid states},\
  }}\href {\doibase 10.1103/RevModPhys.89.025003} {\bibfield  {journal}
  {\bibinfo  {journal} {Reviews of Modern Physics}\ }\textbf {\bibinfo {volume}
  {89}},\ \bibinfo {pages} {025003} (\bibinfo {year} {2017})}\BibitemShut
  {NoStop}%
\bibitem [{\citenamefont {Broholm}\ \emph {et~al.}(2020)\citenamefont
  {Broholm}, \citenamefont {Cava}, \citenamefont {Kivelson}, \citenamefont
  {Nocera}, \citenamefont {Norman},\ and\ \citenamefont
  {Senthil}}]{broholm_quantum_2020}%
  \BibitemOpen
  \bibfield  {author} {\bibinfo {author} {\bibfnamefont {C.}~\bibnamefont
  {Broholm}}, \bibinfo {author} {\bibfnamefont {R.~J.}\ \bibnamefont {Cava}},
  \bibinfo {author} {\bibfnamefont {S.~A.}\ \bibnamefont {Kivelson}}, \bibinfo
  {author} {\bibfnamefont {D.~G.}\ \bibnamefont {Nocera}}, \bibinfo {author}
  {\bibfnamefont {M.~R.}\ \bibnamefont {Norman}}, \ and\ \bibinfo {author}
  {\bibfnamefont {T.}~\bibnamefont {Senthil}},\ }\bibfield  {title}
  {{\color{Gray}\small \bibinfo {title} {{Quantum spin liquids}},\ }}\href
  {https://science.sciencemag.org/content/367/6475/eaay0668} {\bibfield
  {journal} {\bibinfo  {journal} {Science}\ }\textbf {\bibinfo {volume} {367}}
  (\bibinfo {year} {2020})}\BibitemShut {NoStop}%
\bibitem [{\citenamefont {Balents}(2010)}]{balents_spin_2010}%
  \BibitemOpen
  \bibfield  {author} {\bibinfo {author} {\bibfnamefont {L.}~\bibnamefont
  {Balents}},\ }\bibfield  {title} {{\color{Gray}\small \bibinfo {title} {Spin
  liquids in frustrated magnets},\ }}\href {\doibase 10.1038/nature08917}
  {\bibfield  {journal} {\bibinfo  {journal} {Nature}\ }\textbf {\bibinfo
  {volume} {464}},\ \bibinfo {pages} {199} (\bibinfo {year}
  {2010})}\BibitemShut {NoStop}%
\bibitem [{\citenamefont {Trebst}(2017)}]{trebst_kitaev_2017}%
  \BibitemOpen
  \bibfield  {author} {\bibinfo {author} {\bibfnamefont {S.}~\bibnamefont
  {Trebst}},\ }\bibfield  {title} {{\color{Gray}\small \bibinfo {title} {Kitaev
  {Materials}},\ }}\href {http://arxiv.org/abs/1701.07056} {\bibfield
  {journal} {\bibinfo  {journal} {arXiv:1701.07056}\ } (\bibinfo {year}
  {2017})}\BibitemShut {NoStop}%
\bibitem [{\citenamefont {Knolle}\ and\ \citenamefont
  {Moessner}(2019)}]{knolle_field_2019}%
  \BibitemOpen
  \bibfield  {author} {\bibinfo {author} {\bibfnamefont {J.}~\bibnamefont
  {Knolle}}\ and\ \bibinfo {author} {\bibfnamefont {R.}~\bibnamefont
  {Moessner}},\ }\bibfield  {title} {{\color{Gray}\small \bibinfo {title} {A
  {Field} {Guide} to {Spin} {Liquids}},\ }}\href {\doibase
  10.1146/annurev-conmatphys-031218-013401} {\bibfield  {journal} {\bibinfo
  {journal} {Annual Review of Condensed Matter Physics}\ }\textbf {\bibinfo
  {volume} {10}},\ \bibinfo {pages} {451} (\bibinfo {year} {2019})}\BibitemShut
  {NoStop}%
\bibitem [{\citenamefont {Stavropoulos}\ \emph {et~al.}(2019)\citenamefont
  {Stavropoulos}, \citenamefont {Pereira},\ and\ \citenamefont
  {Kee}}]{Stavropoulos_S1}%
  \BibitemOpen
  \bibfield  {author} {\bibinfo {author} {\bibfnamefont {P.~P.}\ \bibnamefont
  {Stavropoulos}}, \bibinfo {author} {\bibfnamefont {D.}~\bibnamefont
  {Pereira}}, \ and\ \bibinfo {author} {\bibfnamefont {H.-Y.}\ \bibnamefont
  {Kee}},\ }\bibfield  {title} {{\color{Gray}\small \bibinfo {title}
  {{Microscopic Mechanism for a Higher-Spin Kitaev Model}},\ }}\href {\doibase
  10.1103/PhysRevLett.123.037203} {\bibfield  {journal} {\bibinfo  {journal}
  {Phys. Rev. Lett.}\ }\textbf {\bibinfo {volume} {123}},\ \bibinfo {pages}
  {037203} (\bibinfo {year} {2019})}\BibitemShut {NoStop}%
\bibitem [{\citenamefont {Kitaev}(2006)}]{Kitaev2006anyons}%
  \BibitemOpen
  \bibfield  {author} {\bibinfo {author} {\bibfnamefont {A.}~\bibnamefont
  {Kitaev}},\ }\bibfield  {title} {{\color{Gray}\small \bibinfo {title}
  {{Anyons in an exactly solved model and beyond}},\ }}\href {\doibase
  10.1016/j.aop.2005.10.005} {\bibfield  {journal} {\bibinfo  {journal} {Ann.
  Phys.}\ }\textbf {\bibinfo {volume} {321}},\ \bibinfo {pages} {2} (\bibinfo
  {year} {2006})}\BibitemShut {NoStop}%
\bibitem [{\citenamefont {Burnell}\ and\ \citenamefont
  {Nayak}(2011)}]{NayakParton2011}%
  \BibitemOpen
  \bibfield  {author} {\bibinfo {author} {\bibfnamefont {F.~J.}\ \bibnamefont
  {Burnell}}\ and\ \bibinfo {author} {\bibfnamefont {C.}~\bibnamefont
  {Nayak}},\ }\bibfield  {title} {{\color{Gray}\small \bibinfo {title} {{SU(2)
  slave fermion solution of the Kitaev honeycomb lattice model}},\ }}\href
  {\doibase 10.1103/PhysRevB.84.125125} {\bibfield  {journal} {\bibinfo
  {journal} {Phys. Rev. B}\ }\textbf {\bibinfo {volume} {84}},\ \bibinfo
  {pages} {125125} (\bibinfo {year} {2011})}\BibitemShut {NoStop}%
\bibitem [{\citenamefont {Jackeli}\ and\ \citenamefont
  {Khaliullin}(2009)}]{Jackeli2009}%
  \BibitemOpen
  \bibfield  {author} {\bibinfo {author} {\bibfnamefont {G.}~\bibnamefont
  {Jackeli}}\ and\ \bibinfo {author} {\bibfnamefont {G.}~\bibnamefont
  {Khaliullin}},\ }\bibfield  {title} {{\color{Gray}\small \bibinfo {title}
  {{Mott Insulators in the Strong Spin-Orbit Coupling Limit: From Heisenberg to
  a Quantum Compass and Kitaev Models}},\ }}\href {\doibase
  10.1103/PhysRevLett.102.017205} {\bibfield  {journal} {\bibinfo  {journal}
  {Phys. Rev. Lett.}\ }\textbf {\bibinfo {volume} {102}},\ \bibinfo {pages}
  {017205} (\bibinfo {year} {2009})}\BibitemShut {NoStop}%
\bibitem [{\citenamefont {Plumb}\ \emph {et~al.}(2014)\citenamefont {Plumb},
  \citenamefont {Clancy}, \citenamefont {Sandilands}, \citenamefont {Shankar},
  \citenamefont {Hu}, \citenamefont {Burch}, \citenamefont {Kee},\ and\
  \citenamefont {Kim}}]{Plumb2014spin}%
  \BibitemOpen
  \bibfield  {author} {\bibinfo {author} {\bibfnamefont {K.~W.}\ \bibnamefont
  {Plumb}}, \bibinfo {author} {\bibfnamefont {J.~P.}\ \bibnamefont {Clancy}},
  \bibinfo {author} {\bibfnamefont {L.~J.}\ \bibnamefont {Sandilands}},
  \bibinfo {author} {\bibfnamefont {V.~V.}\ \bibnamefont {Shankar}}, \bibinfo
  {author} {\bibfnamefont {Y.~F.}\ \bibnamefont {Hu}}, \bibinfo {author}
  {\bibfnamefont {K.~S.}\ \bibnamefont {Burch}}, \bibinfo {author}
  {\bibfnamefont {H.-Y.}\ \bibnamefont {Kee}}, \ and\ \bibinfo {author}
  {\bibfnamefont {Y.-J.}\ \bibnamefont {Kim}},\ }\bibfield  {title}
  {{\color{Gray}\small \bibinfo {title}
  {{$\ensuremath{\alpha}\ensuremath{-}{\mathrm{RuCl}}_{3}$: A spin-orbit
  assisted Mott insulator on a honeycomb lattice}},\ }}\href {\doibase
  10.1103/PhysRevB.90.041112} {\bibfield  {journal} {\bibinfo  {journal} {Phys.
  Rev. B}\ }\textbf {\bibinfo {volume} {90}},\ \bibinfo {pages} {041112}
  (\bibinfo {year} {2014})}\BibitemShut {NoStop}%
\bibitem [{\citenamefont {Banerjee}\ \emph {et~al.}(2016)\citenamefont
  {Banerjee}, \citenamefont {Bridges}, \citenamefont {Yan}, \citenamefont
  {Aczel}, \citenamefont {Li}, \citenamefont {Stone}, \citenamefont {Granroth},
  \citenamefont {Lumsden}, \citenamefont {Yiu}, \citenamefont {Knolle},
  \citenamefont {Bhattacharjee}, \citenamefont {Kovrizhin}, \citenamefont
  {Moessner}, \citenamefont {Tennant}, \citenamefont {Mandrus},\ and\
  \citenamefont {Nagler}}]{Banerjee2016proximate}%
  \BibitemOpen
  \bibfield  {author} {\bibinfo {author} {\bibfnamefont {A.}~\bibnamefont
  {Banerjee}}, \bibinfo {author} {\bibfnamefont {C.~A.}\ \bibnamefont
  {Bridges}}, \bibinfo {author} {\bibfnamefont {J.-Q.}\ \bibnamefont {Yan}},
  \bibinfo {author} {\bibfnamefont {A.~A.}\ \bibnamefont {Aczel}}, \bibinfo
  {author} {\bibfnamefont {L.}~\bibnamefont {Li}}, \bibinfo {author}
  {\bibfnamefont {M.~B.}\ \bibnamefont {Stone}}, \bibinfo {author}
  {\bibfnamefont {G.~E.}\ \bibnamefont {Granroth}}, \bibinfo {author}
  {\bibfnamefont {M.~D.}\ \bibnamefont {Lumsden}}, \bibinfo {author}
  {\bibfnamefont {Y.}~\bibnamefont {Yiu}}, \bibinfo {author} {\bibfnamefont
  {J.}~\bibnamefont {Knolle}}, \bibinfo {author} {\bibfnamefont
  {S.}~\bibnamefont {Bhattacharjee}}, \bibinfo {author} {\bibfnamefont {D.~L.}\
  \bibnamefont {Kovrizhin}}, \bibinfo {author} {\bibfnamefont {R.}~\bibnamefont
  {Moessner}}, \bibinfo {author} {\bibfnamefont {D.~A.}\ \bibnamefont
  {Tennant}}, \bibinfo {author} {\bibfnamefont {D.~G.}\ \bibnamefont
  {Mandrus}}, \ and\ \bibinfo {author} {\bibfnamefont {S.~E.}\ \bibnamefont
  {Nagler}},\ }\bibfield  {title} {{\color{Gray}\small \bibinfo {title}
  {{Proximate Kitaev quantum spin liquid behaviour in a honeycomb magnet}},\
  }}\href {\doibase 10.1038/nmat4604} {\bibfield  {journal} {\bibinfo
  {journal} {Nature Materials}\ }\textbf {\bibinfo {volume} {15}},\ \bibinfo
  {pages} {733} (\bibinfo {year} {2016})}\BibitemShut {NoStop}%
\bibitem [{\citenamefont {Banerjee}\ \emph {et~al.}(2017)\citenamefont
  {Banerjee}, \citenamefont {Yan}, \citenamefont {Knolle}, \citenamefont
  {Bridges}, \citenamefont {Stone}, \citenamefont {Lumsden}, \citenamefont
  {Mandrus}, \citenamefont {Tennant}, \citenamefont {Moessner},\ and\
  \citenamefont {Nagler}}]{Banerjee2016neutron}%
  \BibitemOpen
  \bibfield  {author} {\bibinfo {author} {\bibfnamefont {A.}~\bibnamefont
  {Banerjee}}, \bibinfo {author} {\bibfnamefont {J.}~\bibnamefont {Yan}},
  \bibinfo {author} {\bibfnamefont {J.}~\bibnamefont {Knolle}}, \bibinfo
  {author} {\bibfnamefont {C.~A.}\ \bibnamefont {Bridges}}, \bibinfo {author}
  {\bibfnamefont {M.~B.}\ \bibnamefont {Stone}}, \bibinfo {author}
  {\bibfnamefont {M.~D.}\ \bibnamefont {Lumsden}}, \bibinfo {author}
  {\bibfnamefont {D.~G.}\ \bibnamefont {Mandrus}}, \bibinfo {author}
  {\bibfnamefont {D.~A.}\ \bibnamefont {Tennant}}, \bibinfo {author}
  {\bibfnamefont {R.}~\bibnamefont {Moessner}}, \ and\ \bibinfo {author}
  {\bibfnamefont {S.~E.}\ \bibnamefont {Nagler}},\ }\bibfield  {title}
  {{\color{Gray}\small \bibinfo {title} {Neutron scattering in the proximate
  quantum spin liquid $\alpha-\mathrm{{RuCl}}_3$},\ }}\href {\doibase
  10.1126/science.aah6015} {\bibfield  {journal} {\bibinfo  {journal}
  {Science}\ }\textbf {\bibinfo {volume} {356}},\ \bibinfo {pages} {1055}
  (\bibinfo {year} {2017})}\BibitemShut {NoStop}%
\bibitem [{\citenamefont {Majumder}\ \emph {et~al.}(2015)\citenamefont
  {Majumder}, \citenamefont {Schmidt}, \citenamefont {Rosner}, \citenamefont
  {Tsirlin}, \citenamefont {Yasuoka},\ and\ \citenamefont
  {Baenitz}}]{BaenitzRuClMF2015}%
  \BibitemOpen
  \bibfield  {author} {\bibinfo {author} {\bibfnamefont {M.}~\bibnamefont
  {Majumder}}, \bibinfo {author} {\bibfnamefont {M.}~\bibnamefont {Schmidt}},
  \bibinfo {author} {\bibfnamefont {H.}~\bibnamefont {Rosner}}, \bibinfo
  {author} {\bibfnamefont {A.~A.}\ \bibnamefont {Tsirlin}}, \bibinfo {author}
  {\bibfnamefont {H.}~\bibnamefont {Yasuoka}}, \ and\ \bibinfo {author}
  {\bibfnamefont {M.}~\bibnamefont {Baenitz}},\ }\bibfield  {title}
  {{\color{Gray}\small \bibinfo {title} {{Anisotropic ${\mathrm{Ru}}^{3+}
  4{d}^{5}$ magnetism in the
  $\ensuremath{\alpha}\ensuremath{-}{\mathrm{RuCl}}_{3}$ honeycomb system:
  Susceptibility, specific heat, and zero-field NMR}},\ }}\href {\doibase
  10.1103/PhysRevB.91.180401} {\bibfield  {journal} {\bibinfo  {journal} {Phys.
  Rev. B}\ }\textbf {\bibinfo {volume} {91}},\ \bibinfo {pages} {180401}
  (\bibinfo {year} {2015})}\BibitemShut {NoStop}%
\bibitem [{\citenamefont {Sears}\ \emph {et~al.}(2015)\citenamefont {Sears},
  \citenamefont {Songvilay}, \citenamefont {Plumb}, \citenamefont {Clancy},
  \citenamefont {Qiu}, \citenamefont {Zhao}, \citenamefont {Parshall},\ and\
  \citenamefont {Kim}}]{Sears_magnetic_order_2015}%
  \BibitemOpen
  \bibfield  {author} {\bibinfo {author} {\bibfnamefont {J.~A.}\ \bibnamefont
  {Sears}}, \bibinfo {author} {\bibfnamefont {M.}~\bibnamefont {Songvilay}},
  \bibinfo {author} {\bibfnamefont {K.~W.}\ \bibnamefont {Plumb}}, \bibinfo
  {author} {\bibfnamefont {J.~P.}\ \bibnamefont {Clancy}}, \bibinfo {author}
  {\bibfnamefont {Y.}~\bibnamefont {Qiu}}, \bibinfo {author} {\bibfnamefont
  {Y.}~\bibnamefont {Zhao}}, \bibinfo {author} {\bibfnamefont {D.}~\bibnamefont
  {Parshall}}, \ and\ \bibinfo {author} {\bibfnamefont {Y.-J.}\ \bibnamefont
  {Kim}},\ }\bibfield  {title} {{\color{Gray}\small \bibinfo {title} {{Magnetic
  order in $\ensuremath{\alpha}\ensuremath{-}{\text{RuCl}}_{3}$: A
  honeycomb-lattice quantum magnet with strong spin-orbit coupling}},\ }}\href
  {\doibase 10.1103/PhysRevB.91.144420} {\bibfield  {journal} {\bibinfo
  {journal} {Phys. Rev. B}\ }\textbf {\bibinfo {volume} {91}},\ \bibinfo
  {pages} {144420} (\bibinfo {year} {2015})}\BibitemShut {NoStop}%
\bibitem [{\citenamefont {Singh}\ and\ \citenamefont
  {Gegenwart}(2010)}]{Singh_antiferromagnetic_mott_2010}%
  \BibitemOpen
  \bibfield  {author} {\bibinfo {author} {\bibfnamefont {Y.}~\bibnamefont
  {Singh}}\ and\ \bibinfo {author} {\bibfnamefont {P.}~\bibnamefont
  {Gegenwart}},\ }\bibfield  {title} {{\color{Gray}\small \bibinfo {title}
  {{Antiferromagnetic Mott insulating state in single crystals of the honeycomb
  lattice material ${\text{Na}}_{2}{\text{IrO}}_{3}$}},\ }}\href {\doibase
  10.1103/PhysRevB.82.064412} {\bibfield  {journal} {\bibinfo  {journal} {Phys.
  Rev. B}\ }\textbf {\bibinfo {volume} {82}},\ \bibinfo {pages} {064412}
  (\bibinfo {year} {2010})}\BibitemShut {NoStop}%
\bibitem [{\citenamefont {Chun}\ \emph {et~al.}(2015)\citenamefont {Chun},
  \citenamefont {Kim}, \citenamefont {Kim}, \citenamefont {Zheng},
  \citenamefont {Stoumpos}, \citenamefont {Malliakas}, \citenamefont
  {Mitchell}, \citenamefont {Mehlawat}, \citenamefont {Singh}, \citenamefont
  {Choi}, \citenamefont {Gog}, \citenamefont {Al-Zein}, \citenamefont {Sala},
  \citenamefont {Krisch}, \citenamefont {Chaloupka}, \citenamefont {Jackeli},
  \citenamefont {Khaliullin},\ and\ \citenamefont {Kim}}]{chun_direct_2015}%
  \BibitemOpen
  \bibfield  {author} {\bibinfo {author} {\bibfnamefont {S.~H.}\ \bibnamefont
  {Chun}}, \bibinfo {author} {\bibfnamefont {J.-W.}\ \bibnamefont {Kim}},
  \bibinfo {author} {\bibfnamefont {J.}~\bibnamefont {Kim}}, \bibinfo {author}
  {\bibfnamefont {H.}~\bibnamefont {Zheng}}, \bibinfo {author} {\bibfnamefont
  {C.~C.}\ \bibnamefont {Stoumpos}}, \bibinfo {author} {\bibfnamefont {C.~D.}\
  \bibnamefont {Malliakas}}, \bibinfo {author} {\bibfnamefont {J.~F.}\
  \bibnamefont {Mitchell}}, \bibinfo {author} {\bibfnamefont {K.}~\bibnamefont
  {Mehlawat}}, \bibinfo {author} {\bibfnamefont {Y.}~\bibnamefont {Singh}},
  \bibinfo {author} {\bibfnamefont {Y.}~\bibnamefont {Choi}}, \bibinfo {author}
  {\bibfnamefont {T.}~\bibnamefont {Gog}}, \bibinfo {author} {\bibfnamefont
  {A.}~\bibnamefont {Al-Zein}}, \bibinfo {author} {\bibfnamefont {M.~M.}\
  \bibnamefont {Sala}}, \bibinfo {author} {\bibfnamefont {M.}~\bibnamefont
  {Krisch}}, \bibinfo {author} {\bibfnamefont {J.}~\bibnamefont {Chaloupka}},
  \bibinfo {author} {\bibfnamefont {G.}~\bibnamefont {Jackeli}}, \bibinfo
  {author} {\bibfnamefont {G.}~\bibnamefont {Khaliullin}}, \ and\ \bibinfo
  {author} {\bibfnamefont {B.~J.}\ \bibnamefont {Kim}},\ }\bibfield  {title}
  {{\color{Gray}\small \bibinfo {title} {{Direct evidence for dominant
  bond-directional interactions in a honeycomb lattice iridate
  {Na}$_2${IrO}$_3$}},\ }}\href {https://www.nature.com/articles/nphys3322}
  {\bibfield  {journal} {\bibinfo  {journal} {Nature Physics}\ }\textbf
  {\bibinfo {volume} {11}},\ \bibinfo {pages} {462} (\bibinfo {year}
  {2015})}\BibitemShut {NoStop}%
\bibitem [{\citenamefont {Revelli}\ \emph {et~al.}(2019)\citenamefont
  {Revelli}, \citenamefont {Sala}, \citenamefont {Monaco}, \citenamefont
  {Hickey}, \citenamefont {Becker}, \citenamefont {Freund}, \citenamefont
  {Jesche}, \citenamefont {Gegenwart}, \citenamefont {Eschmann}, \citenamefont
  {Buessen}, \citenamefont {Trebst}, \citenamefont {van Loosdrecht},
  \citenamefont {van~den Brink},\ and\ \citenamefont
  {Gr\"uninger}}]{revelli2019fingerprints}%
  \BibitemOpen
  \bibfield  {author} {\bibinfo {author} {\bibfnamefont {A.}~\bibnamefont
  {Revelli}}, \bibinfo {author} {\bibfnamefont {M.~M.}\ \bibnamefont {Sala}},
  \bibinfo {author} {\bibfnamefont {G.}~\bibnamefont {Monaco}}, \bibinfo
  {author} {\bibfnamefont {C.}~\bibnamefont {Hickey}}, \bibinfo {author}
  {\bibfnamefont {P.}~\bibnamefont {Becker}}, \bibinfo {author} {\bibfnamefont
  {F.}~\bibnamefont {Freund}}, \bibinfo {author} {\bibfnamefont
  {A.}~\bibnamefont {Jesche}}, \bibinfo {author} {\bibfnamefont
  {P.}~\bibnamefont {Gegenwart}}, \bibinfo {author} {\bibfnamefont
  {T.}~\bibnamefont {Eschmann}}, \bibinfo {author} {\bibfnamefont {F.~L.}\
  \bibnamefont {Buessen}}, \bibinfo {author} {\bibfnamefont {S.}~\bibnamefont
  {Trebst}}, \bibinfo {author} {\bibfnamefont {P.~H.~M.}\ \bibnamefont {van
  Loosdrecht}}, \bibinfo {author} {\bibfnamefont {J.}~\bibnamefont {van~den
  Brink}}, \ and\ \bibinfo {author} {\bibfnamefont {M.}~\bibnamefont
  {Gr\"uninger}},\ }\bibfield  {title} {{\color{Gray}\small \bibinfo {title}
  {{Fingerprints of Kitaev physics in the magnetic excitations of honeycomb
  iridates}},\ }}\href {http://arxiv.org/abs/1905.13590} {\bibfield  {journal}
  {\bibinfo  {journal} {arXiv:1905.13590}\ } (\bibinfo {year}
  {2019})}\BibitemShut {NoStop}%
\bibitem [{\citenamefont {Winter}\ \emph {et~al.}(2016)\citenamefont {Winter},
  \citenamefont {Li}, \citenamefont {Jeschke},\ and\ \citenamefont
  {Valent\'{\i}}}]{WinterValenti2016}%
  \BibitemOpen
  \bibfield  {author} {\bibinfo {author} {\bibfnamefont {S.~M.}\ \bibnamefont
  {Winter}}, \bibinfo {author} {\bibfnamefont {Y.}~\bibnamefont {Li}}, \bibinfo
  {author} {\bibfnamefont {H.~O.}\ \bibnamefont {Jeschke}}, \ and\ \bibinfo
  {author} {\bibfnamefont {R.}~\bibnamefont {Valent\'{\i}}},\ }\bibfield
  {title} {{\color{Gray}\small \bibinfo {title} {{Challenges in design of
  Kitaev materials: Magnetic interactions from competing energy scales}},\
  }}\href {\doibase 10.1103/PhysRevB.93.214431} {\bibfield  {journal} {\bibinfo
   {journal} {Phys. Rev. B}\ }\textbf {\bibinfo {volume} {93}},\ \bibinfo
  {pages} {214431} (\bibinfo {year} {2016})}\BibitemShut {NoStop}%
\bibitem [{\citenamefont {Winter}\ \emph {et~al.}(2017)\citenamefont {Winter},
  \citenamefont {Tsirlin}, \citenamefont {Daghofer}, \citenamefont {van~den
  Brink}, \citenamefont {Singh}, \citenamefont {Gegenwart},\ and\ \citenamefont
  {Valent{\'{\i}}}}]{Winter_2017}%
  \BibitemOpen
  \bibfield  {author} {\bibinfo {author} {\bibfnamefont {S.~M.}\ \bibnamefont
  {Winter}}, \bibinfo {author} {\bibfnamefont {A.~A.}\ \bibnamefont {Tsirlin}},
  \bibinfo {author} {\bibfnamefont {M.}~\bibnamefont {Daghofer}}, \bibinfo
  {author} {\bibfnamefont {J.}~\bibnamefont {van~den Brink}}, \bibinfo {author}
  {\bibfnamefont {Y.}~\bibnamefont {Singh}}, \bibinfo {author} {\bibfnamefont
  {P.}~\bibnamefont {Gegenwart}}, \ and\ \bibinfo {author} {\bibfnamefont
  {R.}~\bibnamefont {Valent{\'{\i}}}},\ }\bibfield  {title}
  {{\color{Gray}\small \bibinfo {title} {{Models and materials for generalized
  Kitaev magnetism}},\ }}\href {\doibase 10.1088/1361-648x/aa8cf5} {\bibfield
  {journal} {\bibinfo  {journal} {Journal of Physics: Condensed Matter}\
  }\textbf {\bibinfo {volume} {29}},\ \bibinfo {pages} {493002} (\bibinfo
  {year} {2017})}\BibitemShut {NoStop}%
\bibitem [{\citenamefont {Nayak}\ \emph {et~al.}(2008)\citenamefont {Nayak},
  \citenamefont {Simon}, \citenamefont {Stern}, \citenamefont {Freedman},\ and\
  \citenamefont {Das~Sarma}}]{Nayak_nonabelian_2008}%
  \BibitemOpen
  \bibfield  {author} {\bibinfo {author} {\bibfnamefont {C.}~\bibnamefont
  {Nayak}}, \bibinfo {author} {\bibfnamefont {S.~H.}\ \bibnamefont {Simon}},
  \bibinfo {author} {\bibfnamefont {A.}~\bibnamefont {Stern}}, \bibinfo
  {author} {\bibfnamefont {M.}~\bibnamefont {Freedman}}, \ and\ \bibinfo
  {author} {\bibfnamefont {S.}~\bibnamefont {Das~Sarma}},\ }\bibfield  {title}
  {{\color{Gray}\small \bibinfo {title} {{Non-Abelian anyons and topological
  quantum computation}},\ }}\href {\doibase 10.1103/RevModPhys.80.1083}
  {\bibfield  {journal} {\bibinfo  {journal} {Rev. Mod. Phys.}\ }\textbf
  {\bibinfo {volume} {80}},\ \bibinfo {pages} {1083} (\bibinfo {year}
  {2008})}\BibitemShut {NoStop}%
\bibitem [{\citenamefont {Jiang}\ \emph {et~al.}(2011)\citenamefont {Jiang},
  \citenamefont {Gu}, \citenamefont {Qi},\ and\ \citenamefont
  {Trebst}}]{Jiang2011possible}%
  \BibitemOpen
  \bibfield  {author} {\bibinfo {author} {\bibfnamefont {H.-C.}\ \bibnamefont
  {Jiang}}, \bibinfo {author} {\bibfnamefont {Z.-C.}\ \bibnamefont {Gu}},
  \bibinfo {author} {\bibfnamefont {X.-L.}\ \bibnamefont {Qi}}, \ and\ \bibinfo
  {author} {\bibfnamefont {S.}~\bibnamefont {Trebst}},\ }\bibfield  {title}
  {{\color{Gray}\small \bibinfo {title} {{Possible proximity of the Mott
  insulating iridate Na${}_{2}$IrO${}_{3}$ to a topological phase: Phase
  diagram of the Heisenberg-Kitaev model in a magnetic field}},\ }}\href
  {\doibase 10.1103/PhysRevB.83.245104} {\bibfield  {journal} {\bibinfo
  {journal} {Phys. Rev. B}\ }\textbf {\bibinfo {volume} {83}},\ \bibinfo
  {pages} {245104} (\bibinfo {year} {2011})}\BibitemShut {NoStop}%
\bibitem [{\citenamefont {Fey}(2013)}]{FeyMSc2013}%
  \BibitemOpen
  \bibfield  {author} {\bibinfo {author} {\bibfnamefont {S.}~\bibnamefont
  {Fey}},\ }\bibfield  {title} {{\color{Gray}\small \bibinfo {title}
  {Field-driven instabilities of the non-abelian topological phase in the
  kitaev model},\ }}\href@noop {} {\bibfield  {journal} {\bibinfo  {journal}
  {MSc Thesis, Technische Universit{\"a}t Dortmund}\ } (\bibinfo {year}
  {2013})}\BibitemShut {NoStop}%
\bibitem [{\citenamefont {Zhu}\ \emph {et~al.}(2018)\citenamefont {Zhu},
  \citenamefont {Kimchi}, \citenamefont {Sheng},\ and\ \citenamefont
  {Fu}}]{ZhuAFMK2017}%
  \BibitemOpen
  \bibfield  {author} {\bibinfo {author} {\bibfnamefont {Z.}~\bibnamefont
  {Zhu}}, \bibinfo {author} {\bibfnamefont {I.}~\bibnamefont {Kimchi}},
  \bibinfo {author} {\bibfnamefont {D.~N.}\ \bibnamefont {Sheng}}, \ and\
  \bibinfo {author} {\bibfnamefont {L.}~\bibnamefont {Fu}},\ }\bibfield
  {title} {{\color{Gray}\small \bibinfo {title} {{Robust non-Abelian spin
  liquid and a possible intermediate phase in the antiferromagnetic Kitaev
  model with magnetic field}},\ }}\href {\doibase 10.1103/PhysRevB.97.241110}
  {\bibfield  {journal} {\bibinfo  {journal} {Phys. Rev. B}\ }\textbf {\bibinfo
  {volume} {97}},\ \bibinfo {pages} {241110} (\bibinfo {year}
  {2018})}\BibitemShut {NoStop}%
\bibitem [{\citenamefont {Hickey}\ and\ \citenamefont
  {Trebst}(2019)}]{Hickey2019}%
  \BibitemOpen
  \bibfield  {author} {\bibinfo {author} {\bibfnamefont {C.}~\bibnamefont
  {Hickey}}\ and\ \bibinfo {author} {\bibfnamefont {S.}~\bibnamefont
  {Trebst}},\ }\bibfield  {title} {{\color{Gray}\small \bibinfo {title}
  {{Emergence of a field-driven U(1) spin liquid in the Kitaev honeycomb
  model}},\ }}\href {https://doi.org/10.1038/s41467-019-08459-9} {\bibfield
  {journal} {\bibinfo  {journal} {Nature Communications}\ }\textbf {\bibinfo
  {volume} {10}},\ \bibinfo {pages} {530} (\bibinfo {year} {2019})}\BibitemShut
  {NoStop}%
\bibitem [{\citenamefont {Gohlke}\ \emph {et~al.}(2018)\citenamefont {Gohlke},
  \citenamefont {Moessner},\ and\ \citenamefont {Pollmann}}]{PollmannAFMK2018}%
  \BibitemOpen
  \bibfield  {author} {\bibinfo {author} {\bibfnamefont {M.}~\bibnamefont
  {Gohlke}}, \bibinfo {author} {\bibfnamefont {R.}~\bibnamefont {Moessner}}, \
  and\ \bibinfo {author} {\bibfnamefont {F.}~\bibnamefont {Pollmann}},\
  }\bibfield  {title} {{\color{Gray}\small \bibinfo {title} {{Dynamical and
  topological properties of the Kitaev model in a [111] magnetic field}},\
  }}\href {\doibase 10.1103/PhysRevB.98.014418} {\bibfield  {journal} {\bibinfo
   {journal} {Phys. Rev. B}\ }\textbf {\bibinfo {volume} {98}},\ \bibinfo
  {pages} {014418} (\bibinfo {year} {2018})}\BibitemShut {NoStop}%
\bibitem [{\citenamefont {Jiang}\ \emph {et~al.}(2018)\citenamefont {Jiang},
  \citenamefont {Wang}, \citenamefont {Huang},\ and\ \citenamefont
  {Lu}}]{jiang2018field}%
  \BibitemOpen
  \bibfield  {author} {\bibinfo {author} {\bibfnamefont {H.-C.}\ \bibnamefont
  {Jiang}}, \bibinfo {author} {\bibfnamefont {C.-Y.}\ \bibnamefont {Wang}},
  \bibinfo {author} {\bibfnamefont {B.}~\bibnamefont {Huang}}, \ and\ \bibinfo
  {author} {\bibfnamefont {Y.-M.}\ \bibnamefont {Lu}},\ }\bibfield  {title}
  {{\color{Gray}\small \bibinfo {title} {Field induced quantum spin liquid with
  spinon fermi surfaces in the kitaev model},\ }}\href
  {http://arxiv.org/abs/1809.08247} {\bibfield  {journal} {\bibinfo  {journal}
  {arXiv:1809.08247}\ } (\bibinfo {year} {2018})}\BibitemShut {NoStop}%
\bibitem [{\citenamefont {Patel}\ and\ \citenamefont
  {Trivedi}(2019)}]{Patel2019}%
  \BibitemOpen
  \bibfield  {author} {\bibinfo {author} {\bibfnamefont {N.~D.}\ \bibnamefont
  {Patel}}\ and\ \bibinfo {author} {\bibfnamefont {N.}~\bibnamefont
  {Trivedi}},\ }\bibfield  {title} {{\color{Gray}\small \bibinfo {title}
  {Magnetic field-induced intermediate quantum spin liquid with a spinon fermi
  surface},\ }}\href {\doibase 10.1073/pnas.1821406116} {\bibfield  {journal}
  {\bibinfo  {journal} {Proceedings of the National Academy of Sciences}\
  }\textbf {\bibinfo {volume} {116}},\ \bibinfo {pages} {12199} (\bibinfo
  {year} {2019})}\BibitemShut {NoStop}%
\bibitem [{\citenamefont {Nasu}\ \emph {et~al.}(2018)\citenamefont {Nasu},
  \citenamefont {Kato}, \citenamefont {Kamiya},\ and\ \citenamefont
  {Motome}}]{Nasu2018}%
  \BibitemOpen
  \bibfield  {author} {\bibinfo {author} {\bibfnamefont {J.}~\bibnamefont
  {Nasu}}, \bibinfo {author} {\bibfnamefont {Y.}~\bibnamefont {Kato}}, \bibinfo
  {author} {\bibfnamefont {Y.}~\bibnamefont {Kamiya}}, \ and\ \bibinfo {author}
  {\bibfnamefont {Y.}~\bibnamefont {Motome}},\ }\bibfield  {title}
  {{\color{Gray}\small \bibinfo {title} {Successive majorana topological
  transitions driven by a magnetic field in the kitaev model},\ }}\href
  {\doibase 10.1103/PhysRevB.98.060416} {\bibfield  {journal} {\bibinfo
  {journal} {Phys. Rev. B}\ }\textbf {\bibinfo {volume} {98}},\ \bibinfo
  {pages} {060416} (\bibinfo {year} {2018})}\BibitemShut {NoStop}%
\bibitem [{\citenamefont {Kaib}\ \emph {et~al.}(2019)\citenamefont {Kaib},
  \citenamefont {Winter},\ and\ \citenamefont {Valent\'{\i}}}]{Kaib2019}%
  \BibitemOpen
  \bibfield  {author} {\bibinfo {author} {\bibfnamefont {D.~A.~S.}\
  \bibnamefont {Kaib}}, \bibinfo {author} {\bibfnamefont {S.~M.}\ \bibnamefont
  {Winter}}, \ and\ \bibinfo {author} {\bibfnamefont {R.}~\bibnamefont
  {Valent\'{\i}}},\ }\bibfield  {title} {{\color{Gray}\small \bibinfo {title}
  {Kitaev honeycomb models in magnetic fields: Dynamical response and dual
  models},\ }}\href {\doibase 10.1103/PhysRevB.100.144445} {\bibfield
  {journal} {\bibinfo  {journal} {Phys. Rev. B}\ }\textbf {\bibinfo {volume}
  {100}},\ \bibinfo {pages} {144445} (\bibinfo {year} {2019})}\BibitemShut
  {NoStop}%
\bibitem [{\citenamefont {Zvereva}\ \emph {et~al.}(2015)\citenamefont
  {Zvereva}, \citenamefont {Stratan}, \citenamefont {Ovchenkov}, \citenamefont
  {Nalbandyan}, \citenamefont {Lin}, \citenamefont {Vavilova}, \citenamefont
  {Iakovleva}, \citenamefont {Abdel-Hafiez}, \citenamefont {Silhanek},
  \citenamefont {Chen}, \citenamefont {Stroppa}, \citenamefont {Picozzi},
  \citenamefont {Jeschke}, \citenamefont {Valent\'{\i}},\ and\ \citenamefont
  {Vasiliev}}]{zvereva_zigzag_2015}%
  \BibitemOpen
  \bibfield  {author} {\bibinfo {author} {\bibfnamefont {E.~A.}\ \bibnamefont
  {Zvereva}}, \bibinfo {author} {\bibfnamefont {M.~I.}\ \bibnamefont
  {Stratan}}, \bibinfo {author} {\bibfnamefont {Y.~A.}\ \bibnamefont
  {Ovchenkov}}, \bibinfo {author} {\bibfnamefont {V.~B.}\ \bibnamefont
  {Nalbandyan}}, \bibinfo {author} {\bibfnamefont {J.-Y.}\ \bibnamefont {Lin}},
  \bibinfo {author} {\bibfnamefont {E.~L.}\ \bibnamefont {Vavilova}}, \bibinfo
  {author} {\bibfnamefont {M.~F.}\ \bibnamefont {Iakovleva}}, \bibinfo {author}
  {\bibfnamefont {M.}~\bibnamefont {Abdel-Hafiez}}, \bibinfo {author}
  {\bibfnamefont {A.~V.}\ \bibnamefont {Silhanek}}, \bibinfo {author}
  {\bibfnamefont {X.-J.}\ \bibnamefont {Chen}}, \bibinfo {author}
  {\bibfnamefont {A.}~\bibnamefont {Stroppa}}, \bibinfo {author} {\bibfnamefont
  {S.}~\bibnamefont {Picozzi}}, \bibinfo {author} {\bibfnamefont {H.~O.}\
  \bibnamefont {Jeschke}}, \bibinfo {author} {\bibfnamefont {R.}~\bibnamefont
  {Valent\'{\i}}}, \ and\ \bibinfo {author} {\bibfnamefont {A.~N.}\
  \bibnamefont {Vasiliev}},\ }\bibfield  {title} {{\color{Gray}\small \bibinfo
  {title} {{Zigzag antiferromagnetic quantum ground state in monoclinic
  honeycomb lattice antimonates $A_3$Ni$_2$SbO$_6$($A$=Li,Na)}},\ }}\href
  {\doibase 10.1103/PhysRevB.92.144401} {\bibfield  {journal} {\bibinfo
  {journal} {Phys. Rev. B}\ }\textbf {\bibinfo {volume} {92}},\ \bibinfo
  {pages} {144401} (\bibinfo {year} {2015})}\BibitemShut {NoStop}%
\bibitem [{\citenamefont {Baskaran}\ \emph {et~al.}(2008)\citenamefont
  {Baskaran}, \citenamefont {Sen},\ and\ \citenamefont
  {Shankar}}]{baskaran_orderdisorder_2008}%
  \BibitemOpen
  \bibfield  {author} {\bibinfo {author} {\bibfnamefont {G.}~\bibnamefont
  {Baskaran}}, \bibinfo {author} {\bibfnamefont {D.}~\bibnamefont {Sen}}, \
  and\ \bibinfo {author} {\bibfnamefont {R.}~\bibnamefont {Shankar}},\
  }\bibfield  {title} {{\color{Gray}\small \bibinfo {title} {{Spin-$S$ Kitaev
  model: Classical ground states, order from disorder, and exact correlation
  functions}},\ }}\href {\doibase 10.1103/PhysRevB.78.115116} {\bibfield
  {journal} {\bibinfo  {journal} {Phys. Rev. B}\ }\textbf {\bibinfo {volume}
  {78}},\ \bibinfo {pages} {115116} (\bibinfo {year} {2008})}\BibitemShut
  {NoStop}%
\bibitem [{\citenamefont {Sandvik}(2010)}]{sandvikreview}%
  \BibitemOpen
  \bibfield  {author} {\bibinfo {author} {\bibfnamefont {A.~W.}\ \bibnamefont
  {Sandvik}},\ }\bibfield  {title} {{\color{Gray}\small \bibinfo {title}
  {Computational studies of quantum spin systems},\ }}\href {\doibase
  10.1063/1.3518900} {\bibfield  {journal} {\bibinfo  {journal} {AIP Conference
  Proceedings}\ }\textbf {\bibinfo {volume} {1297}},\ \bibinfo {pages} {135}
  (\bibinfo {year} {2010})}\BibitemShut {NoStop}%
\bibitem [{\citenamefont {Lehoucq}\ \emph {et~al.}(1997)\citenamefont
  {Lehoucq}, \citenamefont {Sorensen},\ and\ \citenamefont
  {Yang}}]{arpackusers}%
  \BibitemOpen
  \bibfield  {author} {\bibinfo {author} {\bibfnamefont {R.~B.}\ \bibnamefont
  {Lehoucq}}, \bibinfo {author} {\bibfnamefont {D.~C.}\ \bibnamefont
  {Sorensen}}, \ and\ \bibinfo {author} {\bibfnamefont {C.}~\bibnamefont
  {Yang}},\ }\bibfield  {title} {{\color{Gray}\small \bibinfo {title} {{ARPACK
  Users Guide: Solution of Large Scale Eigenvalue Problems by Implicitly
  Restarted Arnoldi Methods}},\ }}\href
  {https://epubs.siam.org/doi/book/10.1137/1.9780898719628} {\bibfield
  {journal} {\bibinfo  {journal} {SIAM}\ } (\bibinfo {year}
  {1997})}\BibitemShut {NoStop}%
\bibitem [{Note1()}]{Note1}%
  \BibitemOpen
  \bibinfo {note} {{Note that we are using a normalized magnetic field vector,
  e.g.~a field along the $[111]$ direction would be $\protect \ensuremath
  {\protect \mathbf {h}}=\left (h,h,h\right )/\protect \sqrt {3}$}}\BibitemShut
  {NoStop}%
\bibitem [{\citenamefont {Koga}\ \emph {et~al.}(2018)\citenamefont {Koga},
  \citenamefont {Tomishige},\ and\ \citenamefont {Nasu}}]{Koga_Spin1}%
  \BibitemOpen
  \bibfield  {author} {\bibinfo {author} {\bibfnamefont {A.}~\bibnamefont
  {Koga}}, \bibinfo {author} {\bibfnamefont {H.}~\bibnamefont {Tomishige}}, \
  and\ \bibinfo {author} {\bibfnamefont {J.}~\bibnamefont {Nasu}},\ }\bibfield
  {title} {{\color{Gray}\small \bibinfo {title} {{Ground-state and
  Thermodynamic Properties of an S = 1 Kitaev Model}},\ }}\href
  {https://doi.org/10.7566/JPSJ.87.063703} {\bibfield  {journal} {\bibinfo
  {journal} {Journal of the Physical Society of Japan}\ }\textbf {\bibinfo
  {volume} {87}},\ \bibinfo {pages} {063703} (\bibinfo {year}
  {2018})}\BibitemShut {NoStop}%
\bibitem [{\citenamefont {Knolle}\ \emph {et~al.}(2014)\citenamefont {Knolle},
  \citenamefont {Kovrizhin}, \citenamefont {Chalker},\ and\ \citenamefont
  {Moessner}}]{Knolle_Dynamics_2014}%
  \BibitemOpen
  \bibfield  {author} {\bibinfo {author} {\bibfnamefont {J.}~\bibnamefont
  {Knolle}}, \bibinfo {author} {\bibfnamefont {D.~L.}\ \bibnamefont
  {Kovrizhin}}, \bibinfo {author} {\bibfnamefont {J.~T.}\ \bibnamefont
  {Chalker}}, \ and\ \bibinfo {author} {\bibfnamefont {R.}~\bibnamefont
  {Moessner}},\ }\bibfield  {title} {{\color{Gray}\small \bibinfo {title}
  {{Dynamics of a Two-Dimensional Quantum Spin Liquid: Signatures of Emergent
  Majorana Fermions and Fluxes}},\ }}\href {\doibase
  10.1103/PhysRevLett.112.207203} {\bibfield  {journal} {\bibinfo  {journal}
  {Phys. Rev. Lett.}\ }\textbf {\bibinfo {volume} {112}},\ \bibinfo {pages}
  {207203} (\bibinfo {year} {2014})}\BibitemShut {NoStop}%
\bibitem [{\citenamefont {Sugiura}\ and\ \citenamefont
  {Shimizu}(2012)}]{TPQS2012}%
  \BibitemOpen
  \bibfield  {author} {\bibinfo {author} {\bibfnamefont {S.}~\bibnamefont
  {Sugiura}}\ and\ \bibinfo {author} {\bibfnamefont {A.}~\bibnamefont
  {Shimizu}},\ }\bibfield  {title} {{\color{Gray}\small \bibinfo {title}
  {{Thermal Pure Quantum States at Finite Temperature}},\ }}\href {\doibase
  10.1103/PhysRevLett.108.240401} {\bibfield  {journal} {\bibinfo  {journal}
  {Phys. Rev. Lett.}\ }\textbf {\bibinfo {volume} {108}},\ \bibinfo {pages}
  {240401} (\bibinfo {year} {2012})}\BibitemShut {NoStop}%
\bibitem [{\citenamefont {Sugiura}\ and\ \citenamefont
  {Shimizu}(2013)}]{CanTPQS2013}%
  \BibitemOpen
  \bibfield  {author} {\bibinfo {author} {\bibfnamefont {S.}~\bibnamefont
  {Sugiura}}\ and\ \bibinfo {author} {\bibfnamefont {A.}~\bibnamefont
  {Shimizu}},\ }\bibfield  {title} {{\color{Gray}\small \bibinfo {title}
  {{Canonical Thermal Pure Quantum State}},\ }}\href {\doibase
  10.1103/PhysRevLett.111.010401} {\bibfield  {journal} {\bibinfo  {journal}
  {Phys. Rev. Lett.}\ }\textbf {\bibinfo {volume} {111}},\ \bibinfo {pages}
  {010401} (\bibinfo {year} {2013})}\BibitemShut {NoStop}%
\bibitem [{\citenamefont {Oitmaa}\ \emph {et~al.}(2018)\citenamefont {Oitmaa},
  \citenamefont {Koga},\ and\ \citenamefont {Singh}}]{oitmaa_2018}%
  \BibitemOpen
  \bibfield  {author} {\bibinfo {author} {\bibfnamefont {J.}~\bibnamefont
  {Oitmaa}}, \bibinfo {author} {\bibfnamefont {A.}~\bibnamefont {Koga}}, \ and\
  \bibinfo {author} {\bibfnamefont {R.~R.~P.}\ \bibnamefont {Singh}},\
  }\bibfield  {title} {{\color{Gray}\small \bibinfo {title} {{Incipient and
  well-developed entropy plateaus in spin-$S$ Kitaev models}},\ }}\href
  {\doibase 10.1103/PhysRevB.98.214404} {\bibfield  {journal} {\bibinfo
  {journal} {Phys. Rev. B}\ }\textbf {\bibinfo {volume} {98}},\ \bibinfo
  {pages} {214404} (\bibinfo {year} {2018})}\BibitemShut {NoStop}%
\bibitem [{\citenamefont {Nasu}\ \emph {et~al.}(2014)\citenamefont {Nasu},
  \citenamefont {Udagawa},\ and\ \citenamefont
  {Motome}}]{Nasu2014vaporization}%
  \BibitemOpen
  \bibfield  {author} {\bibinfo {author} {\bibfnamefont {J.}~\bibnamefont
  {Nasu}}, \bibinfo {author} {\bibfnamefont {M.}~\bibnamefont {Udagawa}}, \
  and\ \bibinfo {author} {\bibfnamefont {Y.}~\bibnamefont {Motome}},\
  }\bibfield  {title} {{\color{Gray}\small \bibinfo {title} {{Vaporization of
  Kitaev Spin Liquids}},\ }}\href {\doibase 10.1103/PhysRevLett.113.197205}
  {\bibfield  {journal} {\bibinfo  {journal} {Phys. Rev. Lett.}\ }\textbf
  {\bibinfo {volume} {113}},\ \bibinfo {pages} {197205} (\bibinfo {year}
  {2014})}\BibitemShut {NoStop}%
\bibitem [{\citenamefont {Chandra}\ \emph {et~al.}(2010)\citenamefont
  {Chandra}, \citenamefont {Ramola},\ and\ \citenamefont {Dhar}}]{Chandra2010}%
  \BibitemOpen
  \bibfield  {author} {\bibinfo {author} {\bibfnamefont {S.}~\bibnamefont
  {Chandra}}, \bibinfo {author} {\bibfnamefont {K.}~\bibnamefont {Ramola}}, \
  and\ \bibinfo {author} {\bibfnamefont {D.}~\bibnamefont {Dhar}},\ }\bibfield
  {title} {{\color{Gray}\small \bibinfo {title} {{Classical Heisenberg spins on
  a hexagonal lattice with Kitaev couplings}},\ }}\href {\doibase
  10.1103/PhysRevE.82.031113} {\bibfield  {journal} {\bibinfo  {journal} {Phys.
  Rev. E}\ }\textbf {\bibinfo {volume} {82}},\ \bibinfo {pages} {031113}
  (\bibinfo {year} {2010})}\BibitemShut {NoStop}%
\bibitem [{\citenamefont {Sela}\ \emph {et~al.}(2014)\citenamefont {Sela},
  \citenamefont {Jiang}, \citenamefont {Gerlach},\ and\ \citenamefont
  {Trebst}}]{Sela2014}%
  \BibitemOpen
  \bibfield  {author} {\bibinfo {author} {\bibfnamefont {E.}~\bibnamefont
  {Sela}}, \bibinfo {author} {\bibfnamefont {H.-C.}\ \bibnamefont {Jiang}},
  \bibinfo {author} {\bibfnamefont {M.~H.}\ \bibnamefont {Gerlach}}, \ and\
  \bibinfo {author} {\bibfnamefont {S.}~\bibnamefont {Trebst}},\ }\bibfield
  {title} {{\color{Gray}\small \bibinfo {title} {{Order-by-disorder and
  spin-orbital liquids in a distorted Heisenberg-Kitaev model}},\ }}\href
  {\doibase 10.1103/PhysRevB.90.035113} {\bibfield  {journal} {\bibinfo
  {journal} {Phys. Rev. B}\ }\textbf {\bibinfo {volume} {90}},\ \bibinfo
  {pages} {035113} (\bibinfo {year} {2014})}\BibitemShut {NoStop}%
\bibitem [{Note2()}]{Note2}%
  \BibitemOpen
  \bibinfo {note} {{At the very least one would expect a QSL with a finite
  gauge group $\protect \mathbb {Z}_N$ (so that gauge excitations remain
  gapped). The possibility of realizing a scenario with $N>2$, due to the
  larger Hilbert space of the higher-spin Kitaev models, is an interesting
  question.}}\BibitemShut {Stop}%
\bibitem [{Note3()}]{Note3}%
  \BibitemOpen
  \bibinfo {note} {{Note that there have also been alternative routes to
  realizing AFM Kitaev interactions proposed, with spin-1/2 moments \cite
  {motome_AFM_fbased,sugita2019} }}\BibitemShut {NoStop}%
\bibitem [{\citenamefont {Zhu}\ \emph {et~al.}(2020)\citenamefont {Zhu},
  \citenamefont {Weng},\ and\ \citenamefont {Sheng}}]{zhu_magnetic_2020}%
  \BibitemOpen
  \bibfield  {author} {\bibinfo {author} {\bibfnamefont {Z.}~\bibnamefont
  {Zhu}}, \bibinfo {author} {\bibfnamefont {Z.-Y.}\ \bibnamefont {Weng}}, \
  and\ \bibinfo {author} {\bibfnamefont {D.~N.}\ \bibnamefont {Sheng}},\
  }\bibfield  {title} {{\color{Gray}\small \bibinfo {title} {{Magnetic {Field}
  {Induced} {Spin} {Liquids} in {S}=1 {Kitaev} {Honeycomb} {Model}}},\ }}\href
  {http://arxiv.org/abs/2001.05054} {\bibfield  {journal} {\bibinfo  {journal}
  {arXiv:2001.05054}\ } (\bibinfo {year} {2020})}\BibitemShut {NoStop}%
\bibitem [{\citenamefont {Khait}\ \emph {et~al.}(2020)\citenamefont {Khait},
  \citenamefont {Stavropoulos}, \citenamefont {Kee},\ and\ \citenamefont
  {Kim}}]{YB_2020}%
  \BibitemOpen
  \bibfield  {author} {\bibinfo {author} {\bibfnamefont {I.}~\bibnamefont
  {Khait}}, \bibinfo {author} {\bibfnamefont {P.~P.}\ \bibnamefont
  {Stavropoulos}}, \bibinfo {author} {\bibfnamefont {H.-Y.}\ \bibnamefont
  {Kee}}, \ and\ \bibinfo {author} {\bibfnamefont {Y.~B.}\ \bibnamefont
  {Kim}},\ }\bibfield  {title} {{\color{Gray}\small \bibinfo {title}
  {Characterizing spin-one {Kitaev} quantum spin liquids},\ }}\href
  {http://arxiv.org/abs/2001.06000} {\bibfield  {journal} {\bibinfo  {journal}
  {arXiv:2001.06000}\ } (\bibinfo {year} {2020})}\BibitemShut {NoStop}%
\bibitem [{\citenamefont {Jang}\ \emph {et~al.}(2019)\citenamefont {Jang},
  \citenamefont {Sano}, \citenamefont {Kato},\ and\ \citenamefont
  {Motome}}]{motome_AFM_fbased}%
  \BibitemOpen
  \bibfield  {author} {\bibinfo {author} {\bibfnamefont {S.-H.}\ \bibnamefont
  {Jang}}, \bibinfo {author} {\bibfnamefont {R.}~\bibnamefont {Sano}}, \bibinfo
  {author} {\bibfnamefont {Y.}~\bibnamefont {Kato}}, \ and\ \bibinfo {author}
  {\bibfnamefont {Y.}~\bibnamefont {Motome}},\ }\bibfield  {title}
  {{\color{Gray}\small \bibinfo {title} {{Antiferromagnetic Kitaev interaction
  in $f$-electron based honeycomb magnets}},\ }}\href {\doibase
  10.1103/PhysRevB.99.241106} {\bibfield  {journal} {\bibinfo  {journal} {Phys.
  Rev. B}\ }\textbf {\bibinfo {volume} {99}},\ \bibinfo {pages} {241106}
  (\bibinfo {year} {2019})}\BibitemShut {NoStop}%
\bibitem [{\citenamefont {{Sugita}}\ \emph {et~al.}(2019)\citenamefont
  {{Sugita}}, \citenamefont {{Kato}},\ and\ \citenamefont
  {{Motome}}}]{sugita2019}%
  \BibitemOpen
  \bibfield  {author} {\bibinfo {author} {\bibfnamefont {Y.}~\bibnamefont
  {{Sugita}}}, \bibinfo {author} {\bibfnamefont {Y.}~\bibnamefont {{Kato}}}, \
  and\ \bibinfo {author} {\bibfnamefont {Y.}~\bibnamefont {{Motome}}},\
  }\bibfield  {title} {{\color{Gray}\small \bibinfo {title} {{Antiferromagnetic
  Kitaev Interactions in Polar Spin-Orbit Mott Insulators}},\ }}\href
  {http://arxiv.org/abs/1905.12139} {\bibfield  {journal} {\bibinfo  {journal}
  {arXiv:1905.12139}\ } (\bibinfo {year} {2019})}\BibitemShut {NoStop}%
\end{thebibliography}%

\end{document}